\documentclass[twocolumn,showpacs,preprintnumbers,amsmath,amssymb,a4paper,superscriptaddress, nofootinbib]{revtex4-1}

\usepackage{amsmath}
\usepackage{graphicx}
\usepackage{dcolumn}
\usepackage{bm}

\usepackage{latexsym}
\usepackage{amssymb}
\usepackage{epstopdf}
\usepackage{eqnarray} 

\usepackage{mathrsfs}
\usepackage[percent]{overpic}

\begin{document}
\title{Influence of tilted magnetic field on excited states of the two-dimensional hydrogen atom: Quantum Chaos}
\author{Eugene A. Koval}
\email[]{e-cov@yandex.ru}
\affiliation{Bogoliubov Laboratory of Theoretical Physics, Joint Institute for Nuclear Research, Dubna, Moscow Region 141980, Russian Federation}
\affiliation{Department of fundamental problems of microworld physics, Dubna University, Dubna, Moscow Region 141980, Russian Federation}

\author{Oksana A. Koval}
\email[]{kov.oksana20@gmail.com}
\affiliation{Bogoliubov Laboratory of Theoretical Physics, Joint Institute for Nuclear Research, Dubna, Moscow Region 141980, Russian Federation}

\date{\today}

\begin{abstract}\label{txt:abstract}
The aim of the current work is the research of the influence of the \textbf{tilted} magnetic field direction on statistical properties of energy levels of a two-dimensional (2D) hydrogen atom and of an exciton in GaAs/Al$_{0.33}$Ga$_{0.67}$As quantum well. It was discovered that the quantum chaos (QC) is initiated with an increasing angle $\alpha$ between the magnetic field direction and the normal to the atomic plane. QC is characterized by the repulsion of levels leading to the eliminating of the shell structure and by changing the spectrum statistical properties. 
The evolution of the spatial distribution of the square of the absolute value of the wave function at an increasing angle $\alpha$ was described. The differences of calculated dependencies of energies for various excited states on the tilt angle at a wide range of the magnetic field strength were obtained. 
\end{abstract}

\maketitle


\section{Introduction}
Two-dimensional (2D) hydrogen atom, besides being interesting as a purely theoretical model~\cite{Soylu_2006,Turbiner_2014,Turbiner_2015}, was also applied to describe the effect of charged impurity in 2D systems~\cite{Chen_1991,Villalba_1996,Soylu_2007} and effective interaction in the exciton electron-hole pair (magnetoexciton), the motion of which is limited by the plane, in semiconductor 2D heterostructures~\cite{Kallin_1984,Portnoi_2002}.

Quantum chaos in the hydrogen atom in a magnetic field was investigated mostly for the three-dimensional (3D) case.  
A number of recent studies (e.g.,~\cite{Gutzwiller_1971, Friedrich_1989, Harada_1983, Gutzwiller_2013}) showed, that
the dynamics of classical 3D hydrogen atom model is gradually changing from the regular to the chaotic behaviour at increasing strength of the external magnetic field. The manifestation of the quantum chaos, that occurs in the 3D hydrogen in a magnetic field, such as the change of spectrum statistical characteristics~\cite{Bohigas_1984} was established in theoretical works~\cite{Delande_1986,Monteiro_1990}. 
In the paper~\cite{Grabowski_1994} the hydrogen atom was investigated for a particular case of confinement in a 2D parabolic quantum wire approximated by an additional oscillator potential along one of axises for a magnetic field directed \textit{only} along the normal to the atomic plane. 

In contrast to above-mentioned papers we investigate the statistical properties of the energy levels of 2D hydrogen and of exciton in GaAs/Al$_{0.33}$Ga$_{0.67}$As quantum well in the \textit{tilted} magnetic field.  To describe these systems we consider two particles that are attracting by Coulomb potential (without additional parabolic confinement) and interacting with the external tilted magnetic field. In our paper besides statistical properties we present the accurate results of numerical calculations of energies of lowest bound states of 2D hydrogen and of exciton in GaAs/Al$_{0.33}$Ga$_{0.67}$As quantum well for a wide range of magnetic field strength and the tilt angle $\alpha$. The evolution of the lowest bound states’ electron density at an increasing angle $\alpha$ is described. 

This paper includes the formulation of the problem and computational method description, presented in Sec. \ref{FormulationOfProblem}. Obtained results for the magnetic field directed along the normal to the atomic plane are described in Sec.\ref{Section:PerpendicularMagneticField}. Sec.\ref{Section:TiltedMagneticField} includes the obtained results for the tilted magnetic field.

\section{FORMULATION OF THE PROBLEM AND
COMPUTATIONAL METHOD}
\label{FormulationOfProblem}

This paper is devoted to the investigation of the dynamics of 2D hydrogen atom bound states in a homogeneous external magnetic field 
${\boldsymbol{\mathrm B}=B 
\sin (\alpha )\boldsymbol{\mathrm i} +B \cos (\alpha )\boldsymbol{\mathrm k}}$, 
tilted to an angle $\alpha$ with respect to the normal to the plane of electron motion. 
The Hamiltonian of the 2D hydrogen atom in a uniform magnetic field $\boldsymbol{ \mathrm B}$ in the polar coordinates $\boldsymbol{\mathrm \rho}=(\rho,\phi)$ has the form \cite{Turbiner_2014}:
\begin{equation}
\label{eq1}
\mathscr{H}=\frac{\left( {\boldsymbol{\mathrm P}-2 \boldsymbol{\mathrm A}_{\rho}} \right)^2}{2(m_1 + m_2) }+\frac{\left( {\boldsymbol{\mathrm p}- (\mu_2-\mu_1) \boldsymbol{\mathrm A}_{\rho}} \right)^2}{2m_r }-\frac{1}{\rho},
\end{equation}
where 
$m_1$~--- mass of proton, $m_2 = m_e$~--- mass of electron (corresponding mass ratios $\mu_1 = \frac{m_1}{m_1+m_2}$, $\mu_2 = \frac{m_2}{m_1+m_2}$), $m_r=\frac{m_1 m_2}{m_1+m_2}$~--- reduced mass of the system; $\boldsymbol{\mathrm P}$~--- total momentum, and $\boldsymbol{\mathrm p}$~--- relative momentum of the system. We used symmetric gauge for the vector-potential ${\boldsymbol{\mathrm A}_{\rho}=\tfrac{1}{2}\left[ {\boldsymbol{\mathrm B}\times \boldsymbol{\mathrm \rho} } \right]}$, where $\boldsymbol{\mathrm \rho}$ ~--- relative coordinate.
All values in the article are given in atomic units (a.u.): $\hbar =m_e =e=1$. 

\begin{figure*}[ht]
\vspace{-0.4cm}
\hspace{-1.4cm}
\hfill
\begin{minipage}[h]{0.3\linewidth}
\begin{overpic}[height=6cm]{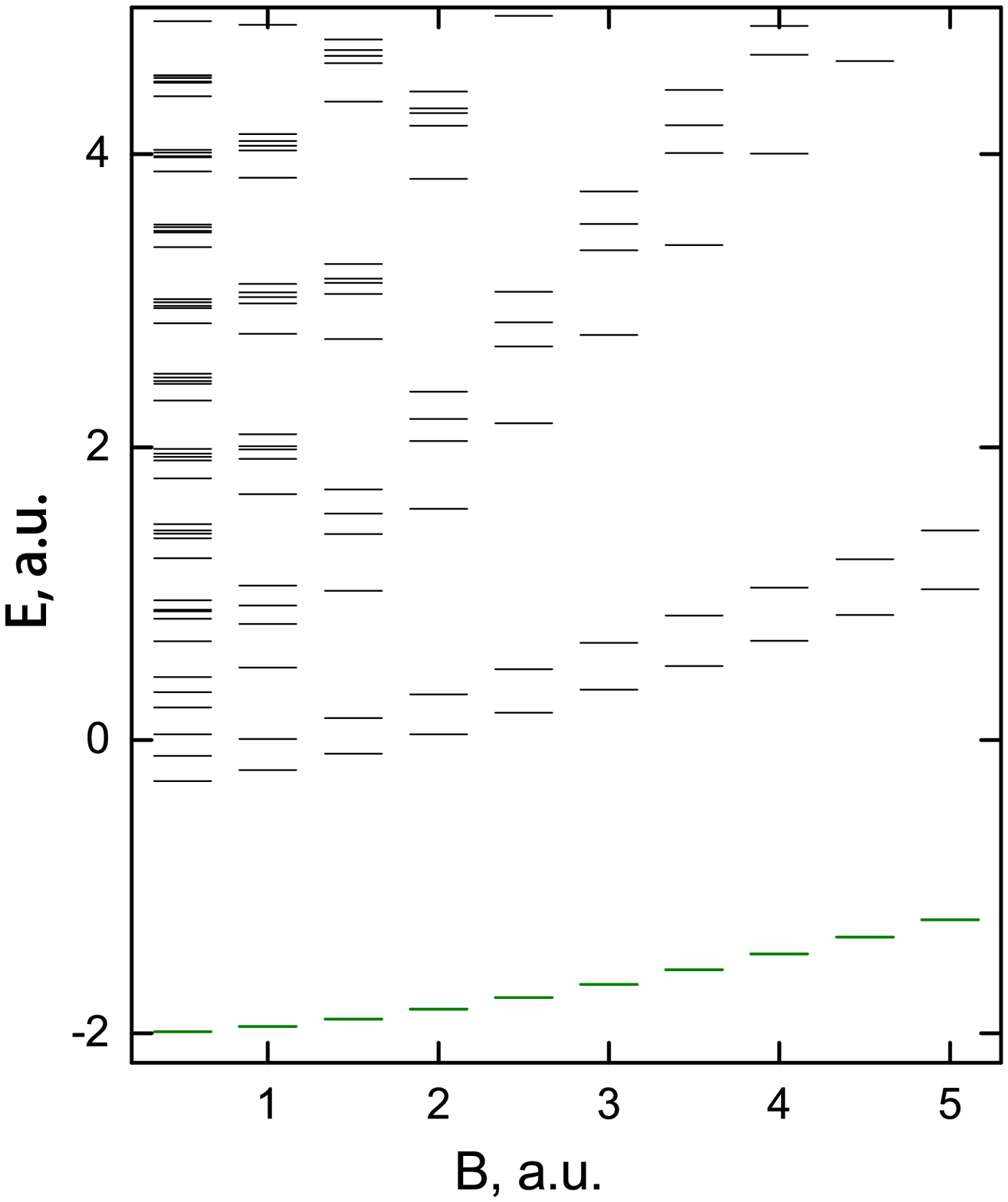}
\put(69,81){\textit{a}}
\end{overpic}
\end{minipage}
\hfill
\begin{minipage}[h]{0.3\linewidth}
\begin{overpic}[height=6cm]{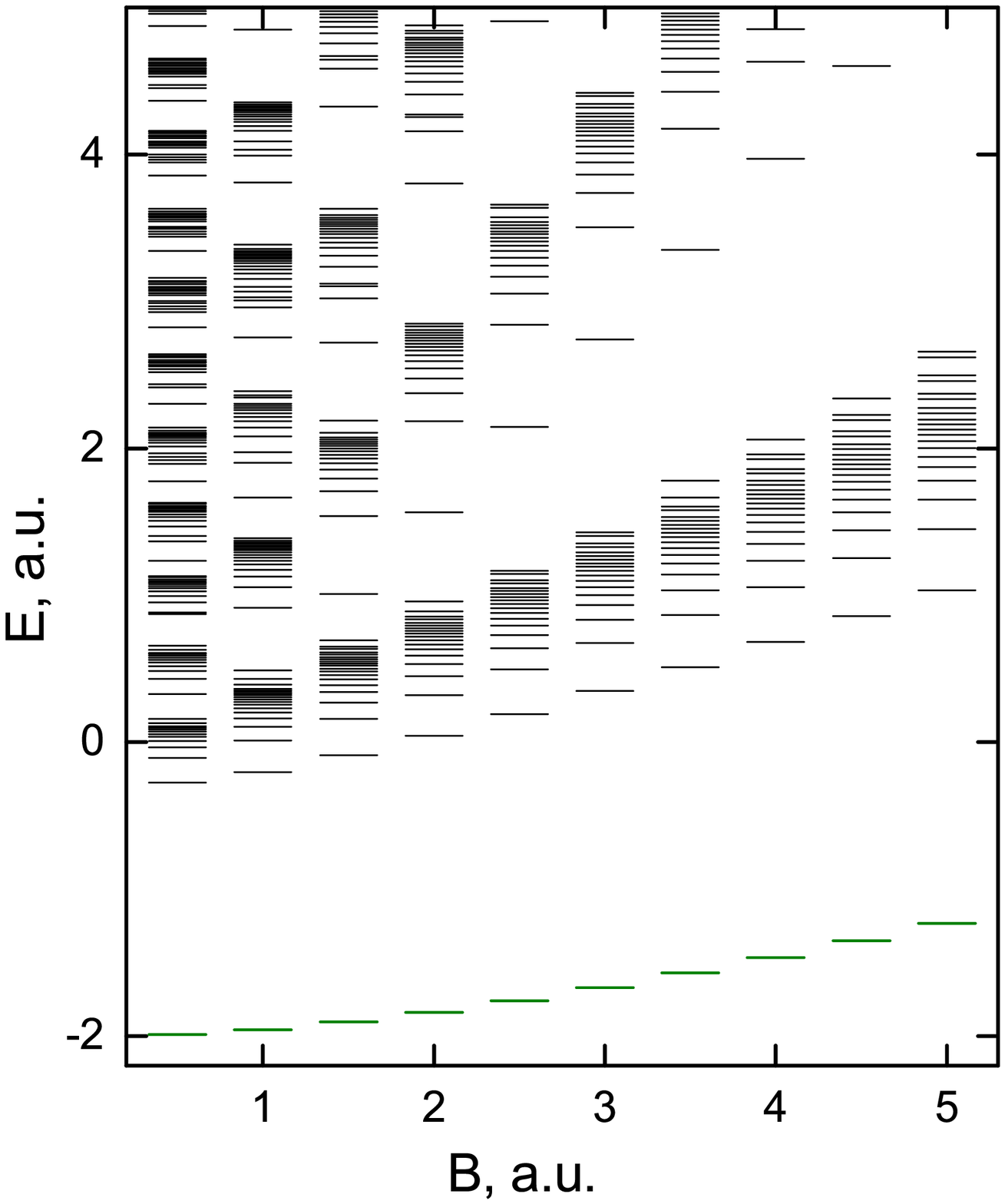}
\put(69,81){\textit{b}}
\end{overpic}
\end{minipage}
\hfill
\begin{minipage}[h]{0.3\linewidth}
\vspace{-0.2cm}
\begin{overpic}[height=6cm]{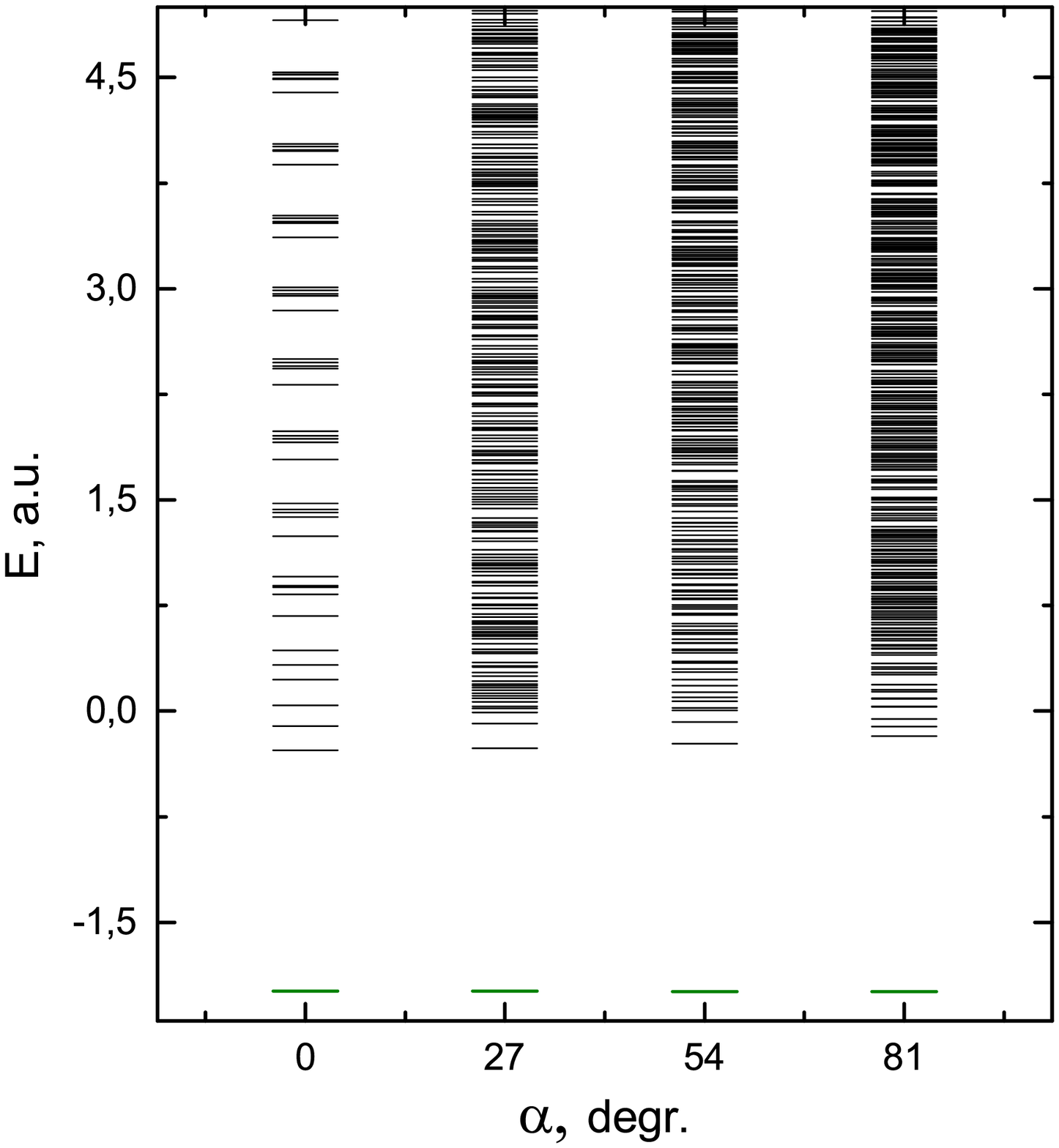}
\put(83,81){\textit{c}}
\end{overpic}
\end{minipage}
\hfill
\caption{(Color online) The calculated spectra of the 2D hydrogen atom at different strengths of the magnetic field: perpendicular to the atomic plane (a), tilted to an angle $\alpha=9^{\circ}$ with respect to the normal to the atomic plane (b), at different angles $\alpha$ and $B = 0.5$~a.u.  (c). Ground state is indicated by green colour.
}
\label{figClusteringAlpha}
\end{figure*}

Similarly to ~\cite{Turbiner_2014}, we considered a system at rest ($\boldsymbol{\mathrm P}=0$): in this case, the motion of the mass center in (\ref{eq1}) is separated from the relative motion. Using the representation of the wave function in the form:
\begin{equation}
\label{eqTurbinerAnsatz}
\Psi(\boldsymbol{\mathrm R},\boldsymbol{\mathrm \rho}) = 
e^{i\boldsymbol{\mathrm  P}\cdot\boldsymbol{\mathrm  R}}\Psi(\boldsymbol{\mathrm \rho}),
\end{equation}
where $\boldsymbol{\mathrm R}$ ~--- coordinate of the mass center, we will obtain the Hamiltonian of relative motion \cite{Turbiner_2014}:
\begin{equation}
\label{eqRelativeMotionMAIN}
H = 
\frac{1}{2m_r} 
 \left[
 {\frac{1}{\rho
 }\frac{\partial }{\partial \rho }\left( {\rho \frac{\partial }{\partial \rho
 }} \right)+\frac{1}{\rho ^2}\frac{\partial ^2}{\partial \phi ^2}} 
\right] + V^{eff}(\rho,\phi), 
\end{equation}
where 
\begin{align}
\label{eqEffectivePotentialLinearTerm}
V^{eff}(\rho,\phi) = \frac{1}{2m_r} 
\left[
\left( \mu_1 - \mu_2 \right) B \cos(\alpha) L_z + 
\right.
\\ 
\label{eqEffectivePotentialQuadraticTerm}
 \left. +  \tfrac{1}{4}{B ^2\rho ^2}\left( {1-\sin ^2(\alpha) \cos ^2 (\phi) } \right) \right] -\frac{1}{\rho}
. 
\end{align}

In order to find the energy levels $E$ and eigenfunctions $\Psi \left( {\rho ,\phi } \right)$ of the Schr{\"o}dinger equation
\begin{equation}
\label{eqBoundStateProblem}
H\Psi({\rho},\phi) =E \Psi({\rho},\phi),
\end{equation}
we use the modification of the discrete variable representation method~\cite{Melezhik_1991}  to reduce the problem to a system of coupled differential equations. To represent the wave function on a uniform grid ${\phi _j =\frac{2\pi j}{2M+1}}(\mbox{where }j=0,1,...,2M)$ by the angular variable $\phi $, we employ eigenfunctions
\begin{equation}
\label{eqKsiDefinition}
\xi _m (\phi )=\frac{1}{\sqrt {2\pi } }e^{im\left( {\phi -\pi } 
\right)}=\frac{(-1)^m}{\sqrt {2\pi } }e^{im\phi },
\end{equation}
of the operator $h^{(0)}(\phi )  \equiv \frac{\partial ^2}{\partial \phi ^2}$ as a Fourier basis. The wave function is sought as the expansion: 
\begin{equation}
\label{eqFullPsiExpansion}
\Psi \left( {\rho ,\phi } \right)=\frac{1}{\sqrt \rho 
} {\sum\limits_{m=-M}^M \sum\limits_{j=0}^{2M}{\xi _m (\phi )\xi 
_{mj}^{-1} \psi _j (\rho )} }, 
\end{equation}
where $\xi _{mj}^{-1} =\frac{2\pi }{2M +1}\xi _{jm}^\ast =\frac{\sqrt{2\pi}
 }{2M +1}e^{-im(\phi _j - \pi) }$ is the inverse matrix to the square matrix $\left({2M +1} \right)\times \left( {2M +1} \right)$ ${\xi _{jm} =\xi _m (\phi _j )}$ determined on the difference grid by the angular variable.
The radial functions $\psi_j(\rho)$ are determined by the values of the wave function on the difference grid  $\phi_j$:
\begin{equation}
\psi_j(\rho)=\sqrt{\rho}\Psi(\rho,\phi_j).
\end{equation}
The Schr{\"o}dinger equation (\ref{eqBoundStateProblem}) in the representation (\ref{eqFullPsiExpansion}) is transformed into a system of $2M+1$ coupled differential equations of the second order:

\begin{multline}
\label{eqShroedEquationInAngularGrid}
\frac{1}{2m_r } 
\left( 
-\frac{\partial^2 }{\partial \rho ^2}\psi_j(\rho)
-\frac{1}{4\rho^2}\psi_j(\rho) +  \sum\limits_{j'=0}^{2M} V^{eff}_{jj'} \psi_{j'}(\rho) -
\right. 
\\
\left. 
- \frac{1}{\rho^2} \sum\limits_{j'=0}^{2M} h^{(0)}_{jj'} \psi_{j'}(\rho)
\right) = E \psi_j(\rho),
\end{multline}
where the potential matrix has the form: 
\begin{multline}
V^{eff}_{jj'}(\rho,\phi) = - \frac{2m_r }{\rho}
\delta_{jj'} + (\mu_p-\mu_e) B \cos(\alpha)h^{(1)}_{jj'} + \\
+ \frac{1}{4}B^2\rho^2\left(1-\sin^2(\alpha)\cos^2(\phi_j)\right)\delta_{jj'},
\end{multline}
and the non-diagonal matrix of the operators $h^{(0)}$ and $h^{(1)}\equiv L_z$ are determined by the following ratios: 
\begin{align}
h^{(0)}_{jj'} = - \sum\limits_{j''=-M}^M j''^2 \xi _{jj''} \xi _{j''j'}^{-1}\\
h^{(1)}_{jj'} = \sum\limits_{j''=-M}^M j'' \xi _{jj''} \xi _{j''j'}^{-1}.
\end{align}

The boundary conditions for the radial functions $\psi_j(\rho)$ are determined by the finiteness of the wave function at zero ($\Psi(\rho,\phi_j)=\frac{\psi_j(\rho)}{\sqrt \rho }\to const$)
\begin{equation}
\label{eqLeftBoundaryCondition}
\psi_j(\rho \to 0)\to const \times {\sqrt \rho} \quad (j=0,1,\ldots ,2M )  
\end{equation}
and by its decreasing at infinity:
\begin{equation}
\label{eqRightBoundaryCondition}
\psi_j(\rho \to \infty) \to 0 \quad (j=0,1,\ldots ,2M ).  
\end{equation}

To solve the eigenvalue problem (\ref{eqShroedEquationInAngularGrid}),(\ref{eqLeftBoundaryCondition}),(\ref{eqRightBoundaryCondition}), the nonuniform grid of the radial variable $\rho$: $\rho_j=\rho_{N} t_{j}^{2}, \quad (j=1,2,\ldots ,N)$ is introduced. Its nodes are determined by mapping $\rho_j \in [0,\rho_{N} \to \infty]$ onto a uniform grid $t_{j}\in [0,1]$.

For discretization we used a finite-difference approximation of the sixth-order accuracy. The eigenvalues of the obtained Hamiltonian matrix are numerically determined by the method of shifted inverse iterations. The algebraic problem arising at each iteration is solved by matrix modification of the sweep algorithm ~\cite{Gelfand_2000} for the block-diagonal matrix.

The applied in this paper numerical method was \textit{successfully verified} in our work~\cite{Koval_2016}, which results are in good agreement with the results of other authors~\cite{Soylu_2006,Turbiner_2014}. The calculated ground state of the exciton in GaAs/Al$_{0.33}$Ga$_{0.67}$As 2D quantum well for $B = 2$ T is equal to $-18,03467$ meV~\cite{Koval_2016} and fully consist with experimental value of the Ref. \cite{Lozovik_2002}.



\section{RESULTS AND DISCUSSION}


\subsection{Magnetic field directed along the normal to the atomic plane}
\label{Section:PerpendicularMagneticField}

In the case of magnetic field directed along the normal to the atomic plane ($\alpha=0^{\circ}$) the Hamiltonian~(\ref{eqRelativeMotionMAIN}) has axial symmetry. It leads to the level clusterization revealed in the calculated spectrum and illustrated in Fig.~\ref{figClusteringAlpha}(a) for different magnetic field strength (from $0.5$ to $5$ a.u.). 

For $B \geq 0.5$ a.u. ($\alpha=0^{\circ}$) 
we observe the atomic level shell structure (see  Fig.~\ref{figClusteringAlpha}(a)), which is formed due to the equidistant spectrum of the quadratic over the magnetic field intensity
``oscillator'' term~(\ref{eqEffectivePotentialQuadraticTerm}) of effective potential $V_{eff}$.

\begin{figure*}[t]
\centering
\hfill
\begin{minipage}[h]{0.3\linewidth}
\begin{overpic}[height=5cm]{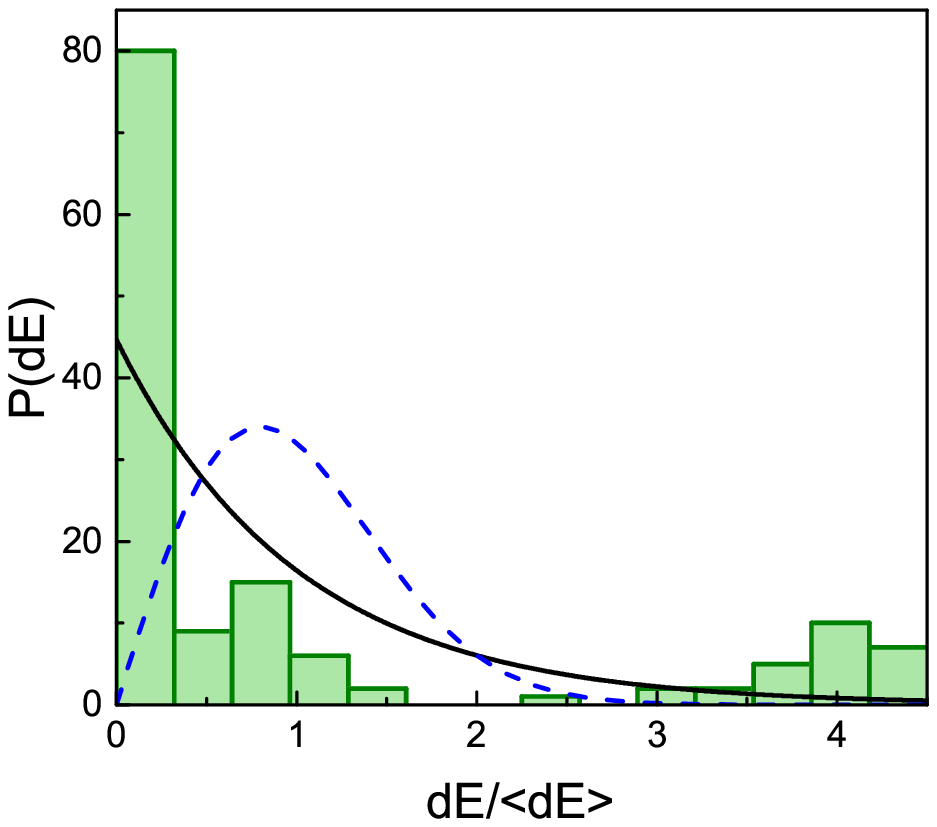}
\put(85,75){\textit{a}}
\end{overpic}
\end{minipage}
\hfill
\begin{minipage}[h]{0.3\linewidth}
\begin{overpic}[height=5cm]{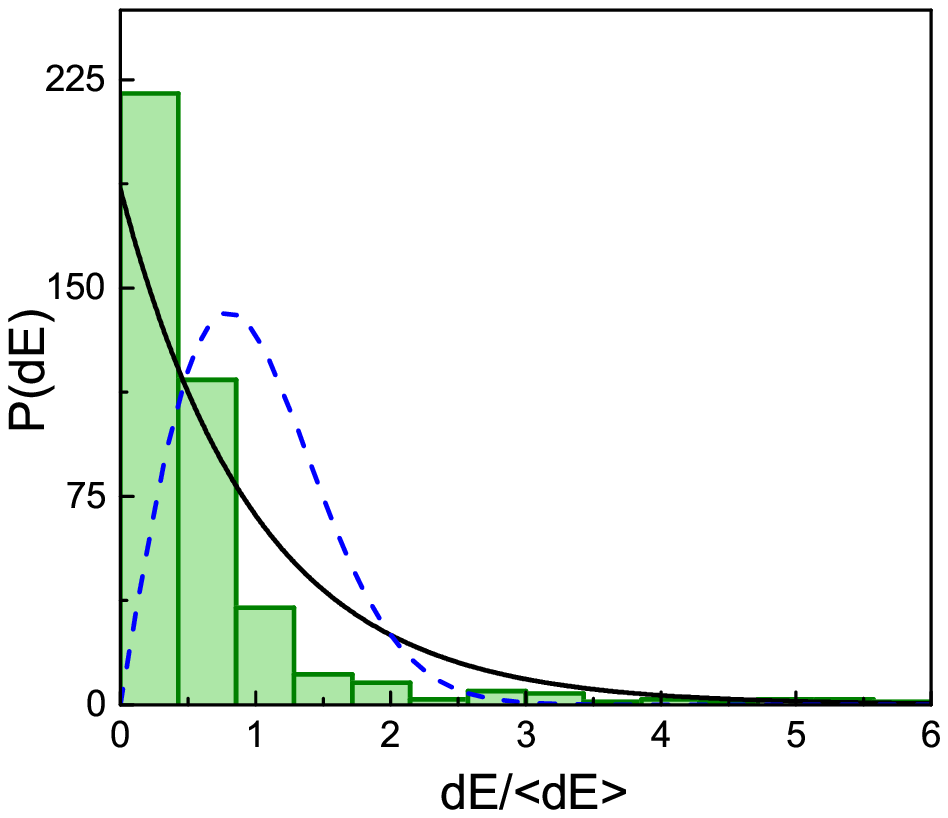}
\put(85,75){\textit{b}}
\end{overpic}
\end{minipage}
\hfill
\begin{minipage}[h]{0.3\linewidth}
\begin{overpic}[height=5cm]{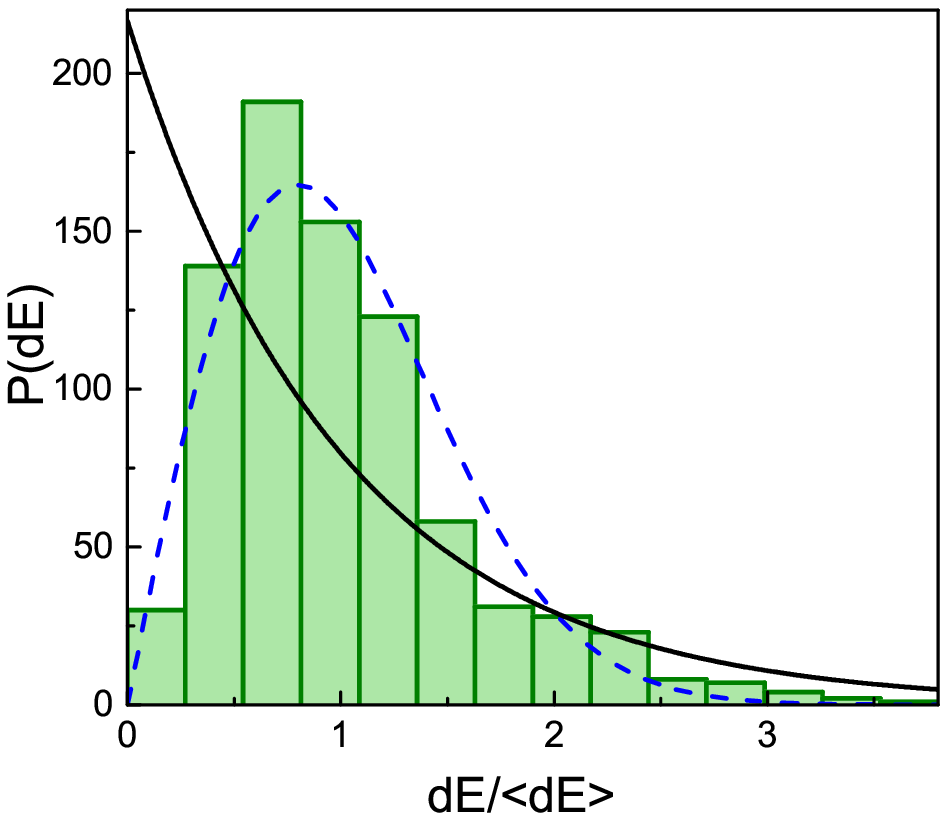}
\put(85,75){\textit{c}}
\end{overpic}
\end{minipage}
\vfill
\begin{minipage}[h]{0.4\linewidth}
\hspace{3cm}
\begin{overpic}[height=5cm]{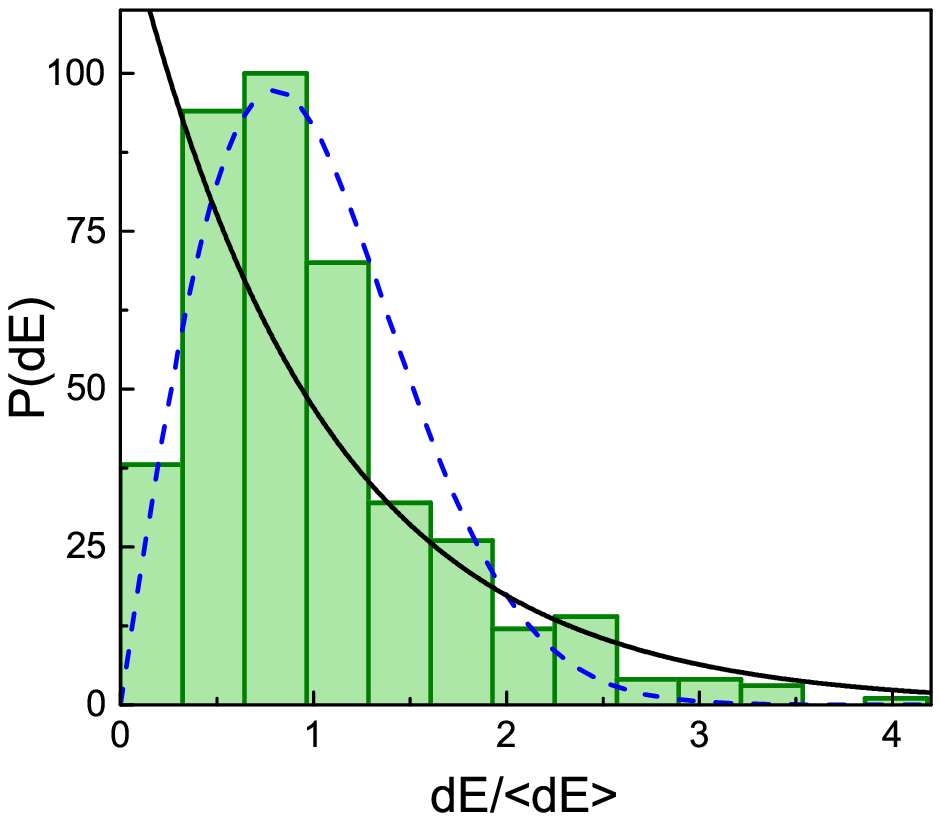}
\put(85,75){\textit{d}}
\end{overpic}
\end{minipage}
\hfill
\begin{minipage}[h]{0.4\linewidth}
\begin{overpic}[height=5cm]{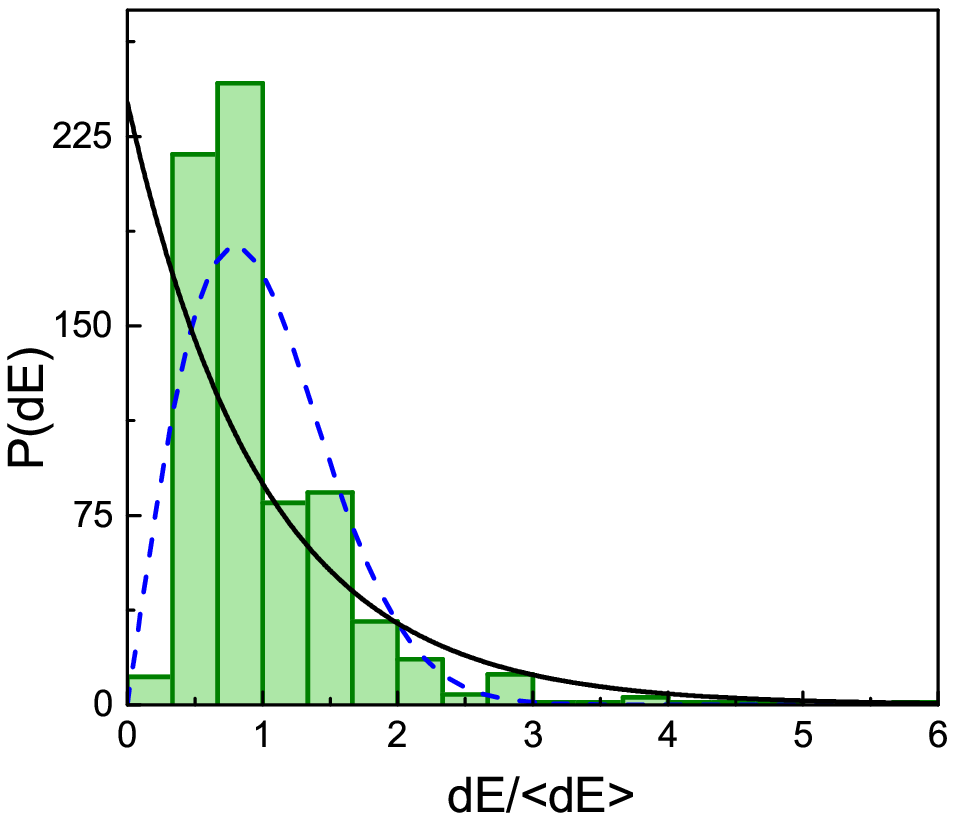}
\put(85,75){\textit{e}}
\end{overpic}
\end{minipage}
\hfill
\caption{
(Color online) The hystograms of unnormalized distributions $P(dE)$ of normalized spacings $dE$/$\left<dE\right>$ at $B=0.5$ a.u. at different tilt angles $\alpha$: $\alpha=0^{\circ}$ (a), $\alpha=9^{\circ}$ (b), $\alpha=27^{\circ}$ (c), $\alpha=54^{\circ}$ (d), $\alpha=81^{\circ}$ (e). Poisson distribution is indicated by solid line, Wigner one~--- by dashed line.
}
 \label{figNNSD}
 \end{figure*}

In order to study behaviour of the system the nearest-neighbor spacing distribution (NNSD) $P(dE)$ were obtained from the calculated 2D hydrogen  spectrum at $B = 0.5$ a.u., where spacings are defined by $dE = E_{j+1}-E_{j}$.
NNSD shown in Fig.~\ref{figNNSD}(a)
is in good agreement with Poisson distribution 
\begin{equation}
P(dE) = \exp(-dE),
\end{equation}
indicating the dominance of the regular motion~\cite{Bohigas_1984}
and level clusterization (the distribution function is maximal for small spacings). NNSD slightly increases at $dE  \approx \omega_L$, where   ${\omega_L \equiv \tfrac{1}{2}B}$ is the Larmor frequency, that
occurs due to the influence of equidistant oscillator~(\ref{eqEffectivePotentialLinearTerm}) spectrum and rather strong considered magnetic fields.

The sizes of the level clusters determined from calculated spectra for $B > 10^3$ a.u.
correspond well to the analytical estimation of the Landau cluster size in strong magnetic fields of Ref.~\cite{Robnik_2003}. 


\subsection{Tilted magnetic fields}
\label{Section:TiltedMagneticField}

In order to investigate the influence of the magnetic field direction on the spectrum of the 2D hydrogen atom we calculate the dependence of the energy of the excited states on the tilted magnetic field strength at different tilt angles $\alpha$ in contrast to Refs.~\cite{Soylu_2006,Turbiner_2014,Robnik_2003}, where the system properties were studied in the magnetic field directed perpendicular to the plane of electron motion.

The absolute values of the energies of the first three excited states numerically calculated in the present work for different strengths and tilt angles of the magnetic field are given in Tables~\ref{tabE_2}--\ref{tabE_4}. It is demonstrated in Tables~\ref{tabE_2}--\ref{tabE_4} that the energy of low-lying excited states with an $\alpha$ increase changes non-linearly by 1.8 times for the first and 4.2 times for the third excited states.

Fig.~\ref{figClusteringAlpha} presents the calculated spectra of the 2D hydrogen atom at different strengths of the magnetic field: perpendicular to the atomic plane (a), tilted to an angle $\alpha=9^{\circ}$ with respect to the normal to the atomic plane (b), at different angles $\alpha$ and $B = 0.5$~a.u. (c). Ground state in Fig.~\ref{figClusteringAlpha} is indicated by green colour.

\begin{figure}[hbtp]
\vspace{-0.4cm}
\centering
\includegraphics[width=0.84 \linewidth]{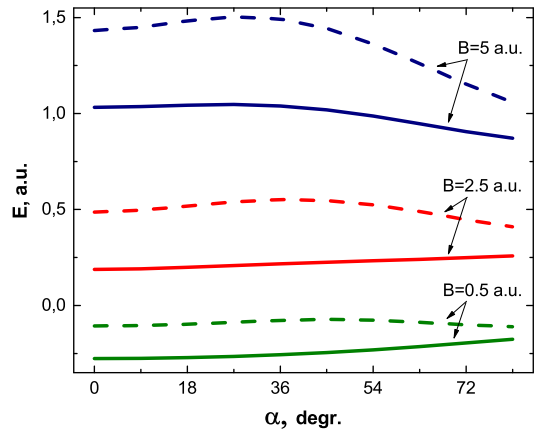}
\caption{
(Color online) Dependencies of energies of the first (solid line) and second (dashed line) excited states of 2D hydrogen atom  on the tilt angle $\alpha$ at different magnetic field strengths ($B = 0.5, 2.5$ and $5.0$ a.u.).
}
\label{figExcitedStatesDependenceOnAlpha}
\end{figure}

Fig.~\ref{figExcitedStatesDependenceOnAlpha} shows the energy dependencies of the first (solid line) and the second (dashed line)
excited states of the 2D hydrogen atom on the tilt angle $\alpha$ at different strength of the magnetic field $B=0.5,2.5$ and $5.0$ a.u.
Analysing Fig.~\ref{figExcitedStatesDependenceOnAlpha}
one can note that the increase of the magnetic field strength leads to change of the character of the energy dependence of the first excited state (solid line) on the tilt angle $\alpha$. At $B<3$ a.u. energy dependence is rising while at $B = 5$ a.u. it turns to non-linear decrease. In the same time the energy dependence of the second excited state always has a maximum in this range of fields.

At small angles ($\alpha \leq 10^{\circ}$), the excited levels split conserving shell structure and Poissonian-like NNSD, which is illustrated in Fig.~\ref{figClusteringAlpha}(b)  and Fig.~\ref{figNNSD}(b).

At further $\alpha$ increasing (e.g. $\alpha \geq 27^{\circ}$ in Fig.~\ref{figNNSD}(c,d,e)) the occurrence of quantum chaos was found, induced by a magnetic field tilting and activation of the anisotropic part of the potential~(\ref{eqEffectivePotentialQuadraticTerm}). It is manifested by the repulsion of adjacent levels~\cite{Haake_2013}. Such energy spectrum behaviour is proved by typical decrease of NNSD for small spacings (see Fig.~\ref{figNNSD}(c,d,e)). Changes in the energy spectrum are accompanied by a corresponding change of NNSD, which is shown in Fig.~\ref{figNNSD}(c,d,e), it also illustrates a good agreement between NNSD and typical for quantum chaos regime Wigner distribution~\cite{Gutzwiller_2013,Bohigas_1984}:
\begin{equation}
P(dE) = \tfrac{1}{2} \pi dE \exp(-\pi dE^2/4).
\end{equation}
 
Analyzing Fig.~\ref{figClusteringAlpha}(c) it should be noted that with increasing anisotropy (by increasing the angle $\alpha$ to $90^{\circ}$) the shell structure, typical for the spectrum of the isotropic case, is eliminated, 
that is related to the splitting of degenerated excited states of the system in the case, when the magnetic field is directed along the normal to the plane of electron motion. The degeneracy due to the axial symmetry occurs as in the Landau states
(in the absence of the Coulomb interaction)~\cite{Robnik_2003}, as in 2D hydrogen atom (in the absence of the magnetic field).

Spatial distributions (SD) of the square of wave function absolute value of the first three excited states significantly transform with the increase of the tilt angle $\alpha$. Due to the anisotropy of the interaction, the projection $L_z$ of angular momentum on the $Z$ axis is not conserved and a preferential direction appears along the projection of the magnetic field on the atomic plane (coinciding with the $X$ axis in the chosen coordinate system). Above-mentioned SD elongate along the $X$ axis with its simultaneous compression along the $Y$ axis, which is strengthened with $\alpha \to 90^{\circ}$.

\begin{figure}[hbtp]
\vspace{-0.4cm}
\centering
\includegraphics[width=0.84 \linewidth]{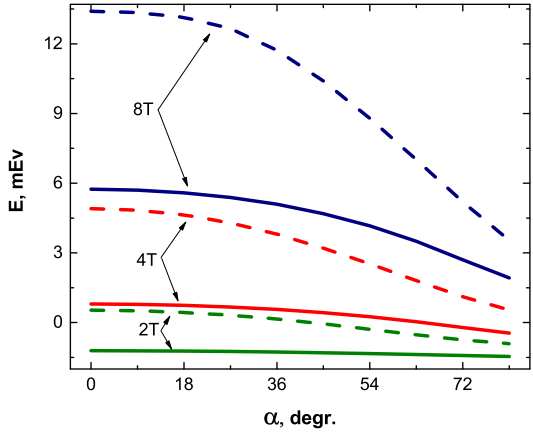}
\caption{
(Color online) Same as in Fig.\ref{figExcitedStatesDependenceOnAlpha} but for the 2D exciton in GaAs/Al$_{0.33}$Ga$_{0.67}$As at different magnetic field strengths: $B=2, 4, 8$ T.
}
\label{figExcitonExcitedStatesDependenceOnAlpha}
\end{figure}
 
 
It should be noted that for the critical case of $\alpha=90^{\circ}$, when the magnetic field lies in the atomic plane $B_z=0$, the oscillator term (3) along the $X$ axis disappears and potential does not lead to the formation of bound states in the positive region of the spectrum. Thus, at $\alpha=90^{\circ}$ energy levels of excited states, rising with an increase of the magnetic field strength, cease to exist at a certain critical (for them) value of the magnetic field. 

In order to consider real physical system such as 2D exciton in GaAs/Al$_{0.33}$Ga$_{0.67}$As we need to 
change the particles masses to effective masses of electron $m_e^*=0.067m_e$ and heavy hole  $m_h=0.18m_e$~\cite{Butov_2001}, and to add $1/ \epsilon$ factor ($\epsilon=12.1$~-- dielectric constant) to Coulomb term. Thus, the Hamiltonian (\ref{eqRelativeMotionMAIN}) for 2D exciton reads 
\begin{equation}
\label{eqRelativeMotionExciton}
H = 
\frac{\boldsymbol{\mathrm p}^2 + 2  (\mu_h-\mu_e) ( \boldsymbol{\mathrm A}_{\rho} \cdot \boldsymbol{\mathrm p})+ \boldsymbol{\mathrm A}_{\rho}^2}{2m_r}-\frac{1}{\epsilon \rho},
\end{equation}
где $\mu_h=\frac{m_h}{m_h+m_e^*}$, $\mu_e=\frac{m_e^*}{m_h+m_e^*}$; $m_r=\frac{m_h m_e^*}{m_h+m_e^*}$. 
Applying our proposed numerical algorithm to the calculation of the 2D exciton spectrum, we obtained the characteristics of the exciton bound states at typical for laboratory experiments magnetic field strength range from $B=2$T to $B=8$T.  
Dependencies of energies of the first (solid line) and second (dashed line) excited states of exciton on the tilt angle $\alpha$ at different magnetic field strengths ($B = 2, 4$ and $8$ T) are presented in Fig.~\ref{figExcitonExcitedStatesDependenceOnAlpha}.  The exciton in GaAs/Al$_{0.33}$Ga$_{0.67}$As first and second excited energies monotonically decrease with the tilt angle $\alpha$ in contrast to those of 2D hydrogen.

We analysed NNSDs for exciton in GaAs/Al$_{0.33}$Ga$_{0.67}$As, which were obtained from the calculated spectrum of energy for different tilt angles and magnetic field strength, and their behaviour is similar to the one of the 2D hydrogen atom. The effect of quantum chaos initiation with the tilt angle $\alpha$ increasing was revealed for exciton in GaAs/Al$_{0.33}$Ga$_{0.67}$As. 

Quantitative differences of 2D hydrogen and exciton energy spectra emerge due to the dielectric constant in Coulomb term and another masses (effective electron and heavy hole masses) of the system particles in the Hamiltonian~(\ref{eqRelativeMotionMAIN}) (e.g. see Ref.~\cite{Butov_2001}).


\section{Conclusion}

The characteristics of the excited states of 2D hydrogen atom and exciton in GaAs/Al$_{0.33}$Ga$_{0.67}$As were investigated in a wide range of magnetic field strength. In the case of the magnetic field directed perpendicular to the atomic plane, level clustering and shell structure in the spectrum are observed. The shell structure at $\alpha=0^{\circ}$ is formed due to equidistant spectrum of a quadratic over magnetic field strength ``oscillator'' term of the effective potential of interaction. Calculated at $\alpha=0^{\circ}$ NNSD reproduce Poisson distribution associated with regular dynamics of the system.
For small tilt angles $\alpha$ the system retains regular behaviour with the splitting of the excited levels due to the anisotropy of the interaction. 
With the increase of the tilt angle $\alpha$ quantum chaos appears in the system, confirmed by the change of NNSD type from the Poisson distribution to the Wigner one. 
The evolution of the spatial distributions of the square of the wave function absolute value with increasing tilt angle $\alpha$ was demonstrated.
We discovered the differences of the calculated energy dependencies  for different excited states on the angle $\alpha$ in a wide range of the magnetic field strength. 

Increasing tilt of the magnetic field direction leads to the complete change of the spectrum of the exciton in GaAs/Al$_{0.33}$Ga$_{0.67}$As. Thus, by changing external magnetic field direction the properties of the semiconductors exciton and impurity absorption spectra can be controlled.

\begin{acknowledgments}
The authors are grateful to V.S. Melezhik and V.V.~Pupishev for discussions of the article. 
The authors acknowledge the support by
the Russian Foundation for Basic Research, Grant No. 16-32-00865.
\end{acknowledgments}

\newpage

\onecolumngrid

 \begin{figure*}[hbt]
\hfill
\begin{minipage}[h]{0.3\linewidth}
\begin{overpic}[height=4cm]{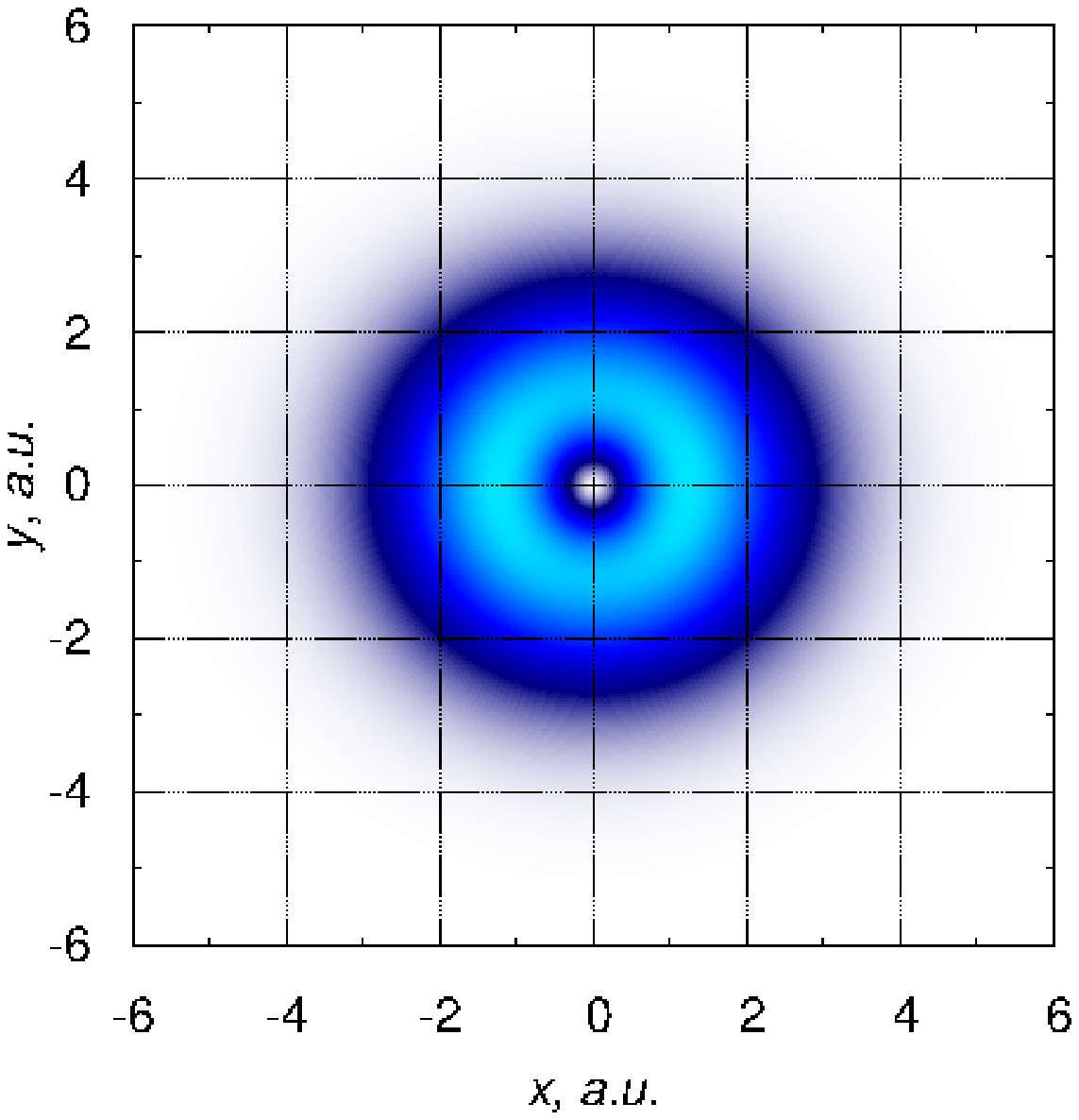}
\put(85,89){\textit{a}}
\end{overpic}
\end{minipage}
\hfill
\begin{minipage}[h]{0.3\linewidth}
\begin{overpic}[height=4cm]{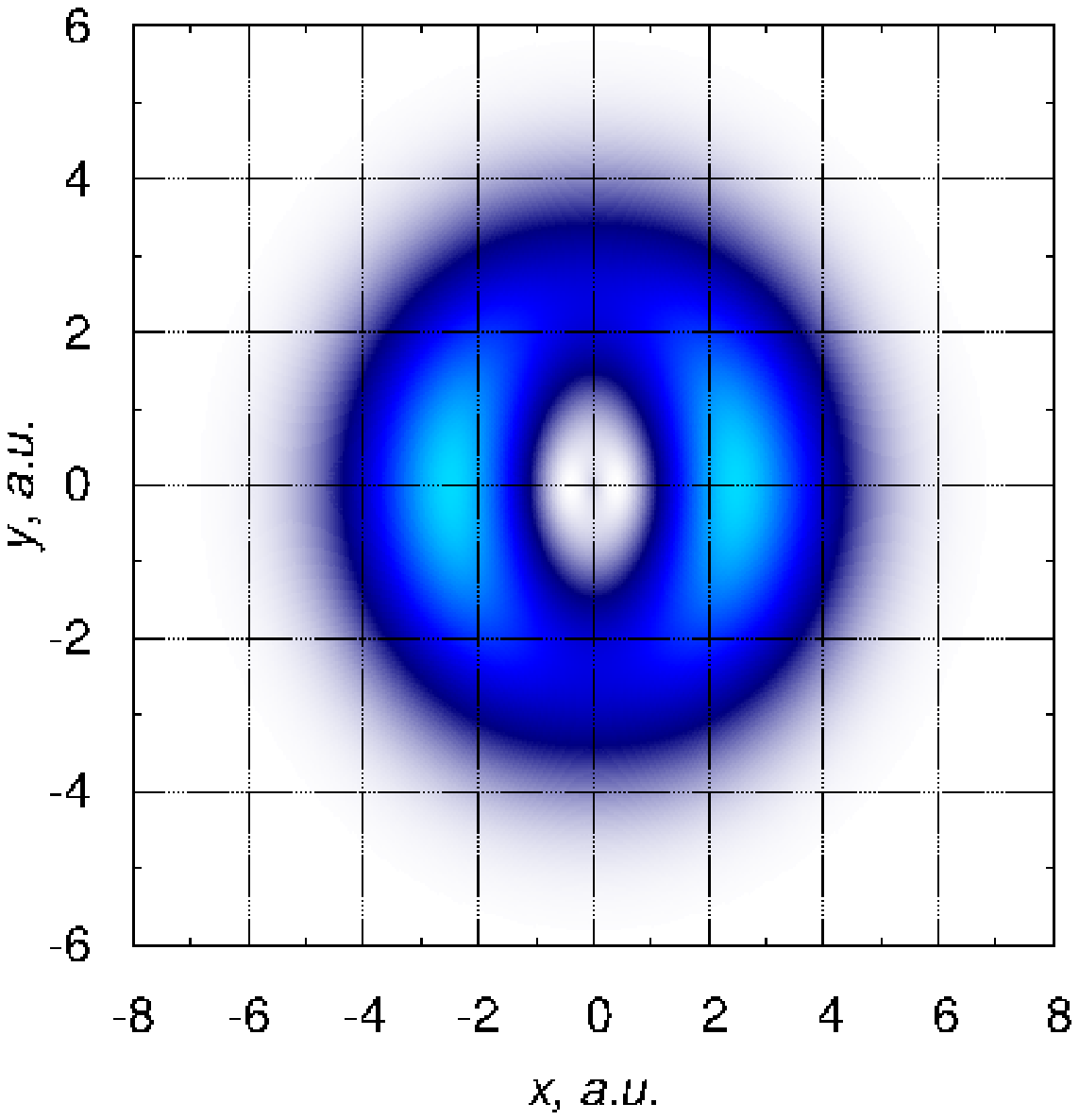}
\put(85,89){\textit{b}}
\end{overpic}
\end{minipage}
\hfill
\begin{minipage}[h]{0.3\linewidth}
\begin{overpic}[height=4cm]{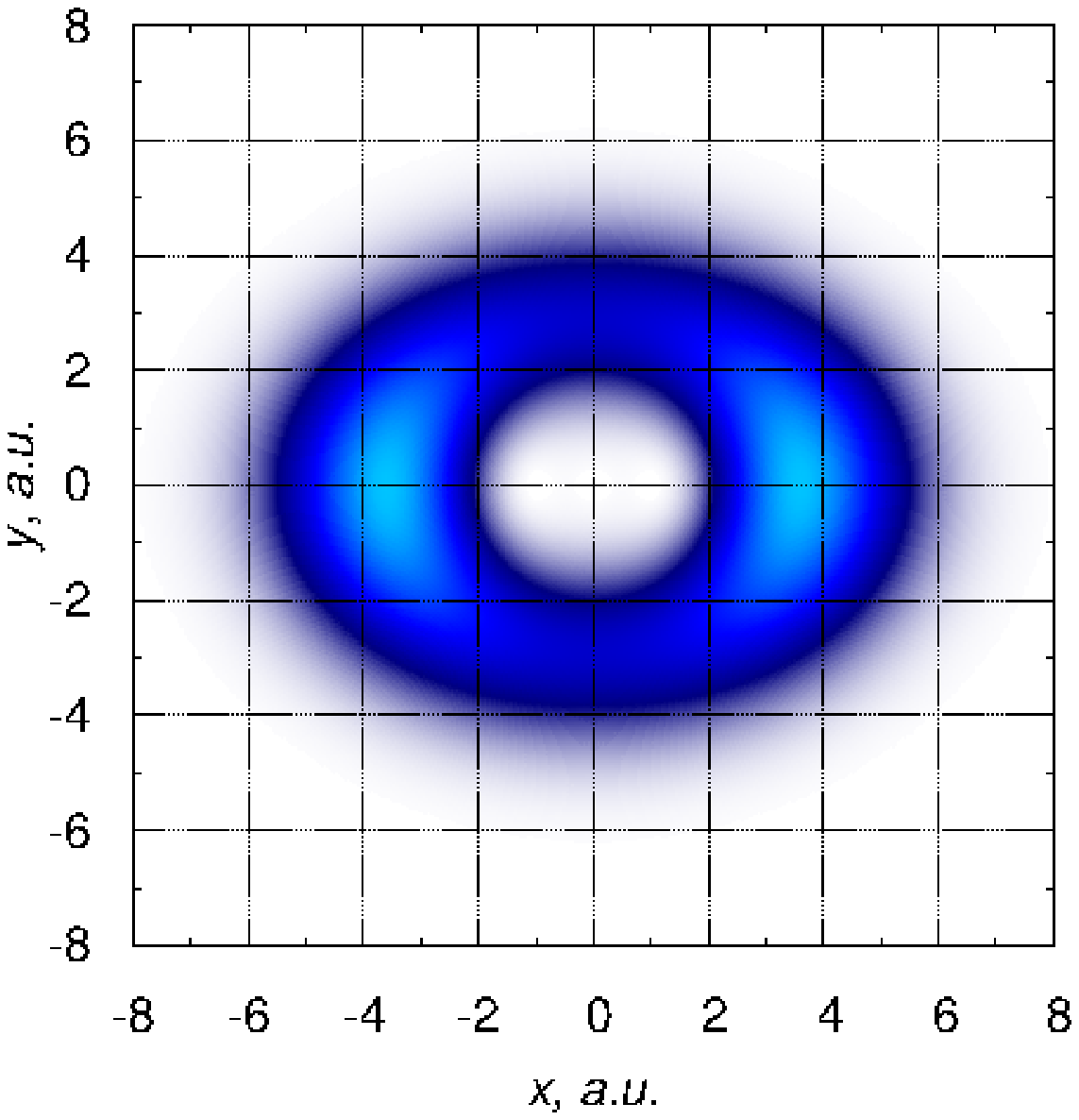}
\put(89,89){\textit{c}}
\end{overpic}
\hfill
\end{minipage}
\vfill
\hfill
\begin{minipage}[h]{0.3\linewidth}
\begin{overpic}[height=4cm]{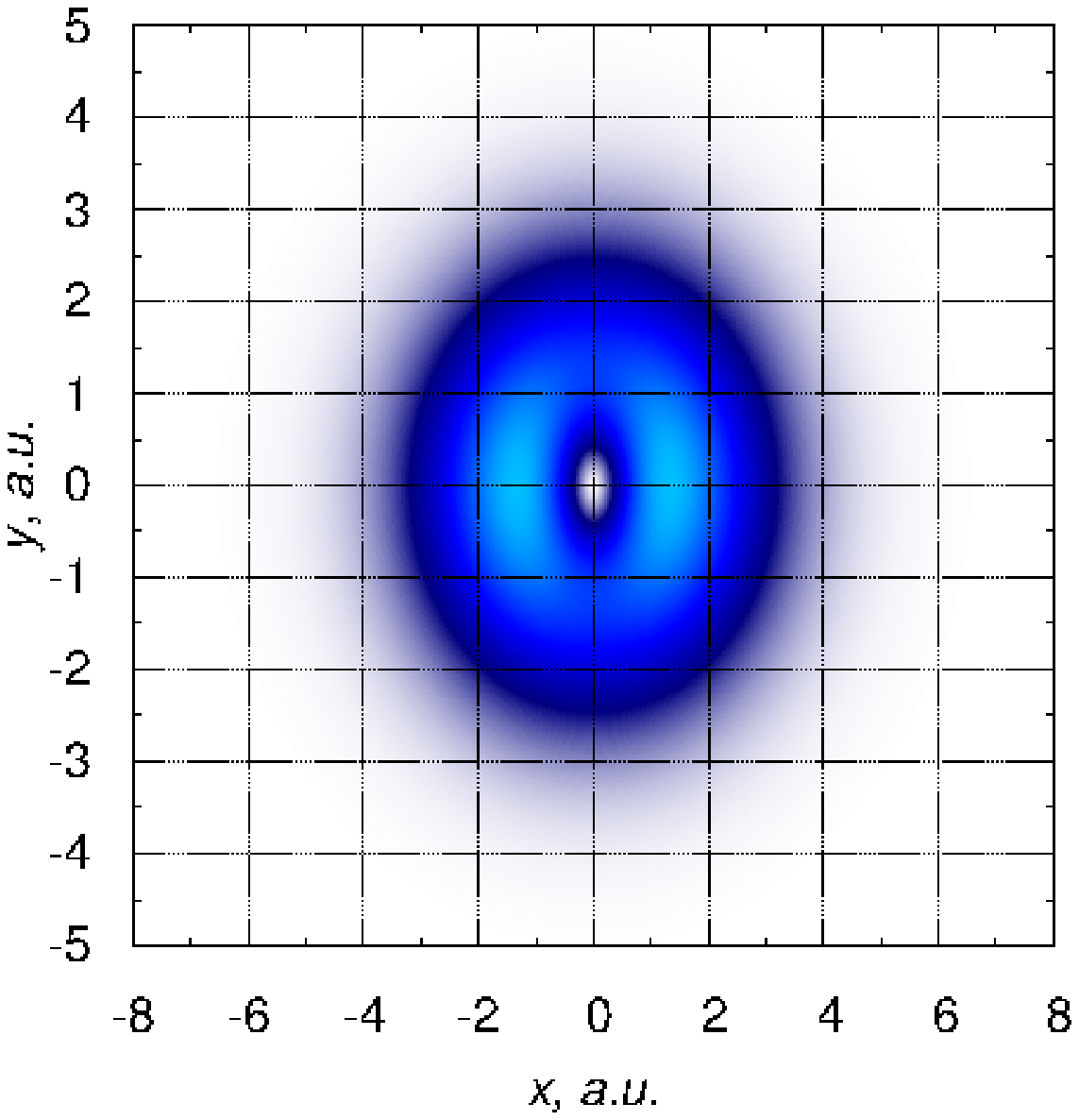}
\put(89,91){\textit{d}}
\end{overpic}
\end{minipage}
\hfill
\begin{minipage}[h]{0.3\linewidth}
\begin{overpic}[height=4cm]{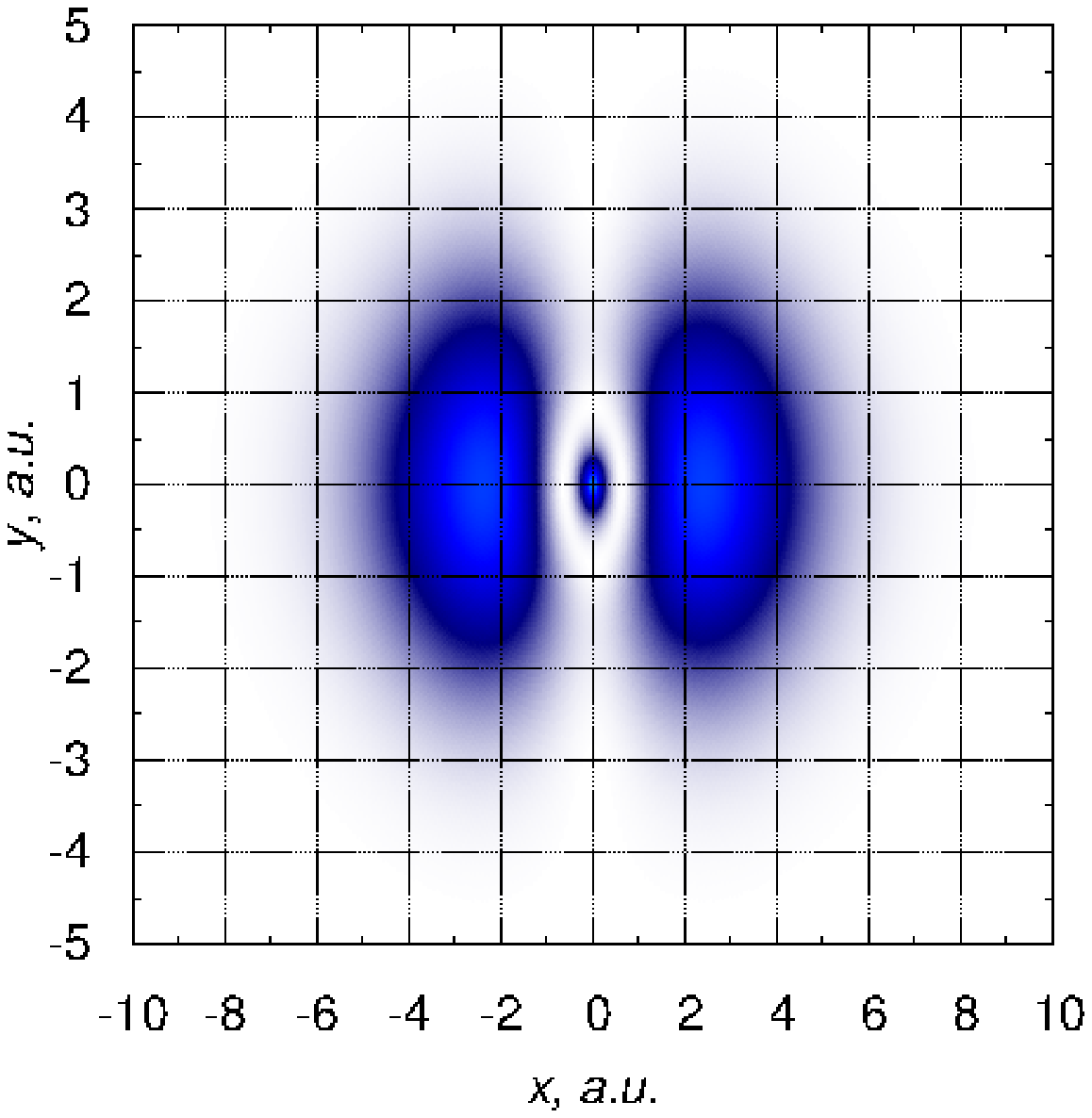}
\put(89,91){\textit{e}}
\end{overpic}
\end{minipage}
\hfill
\begin{minipage}[h]{0.3\linewidth}
\begin{overpic}[height=4cm]{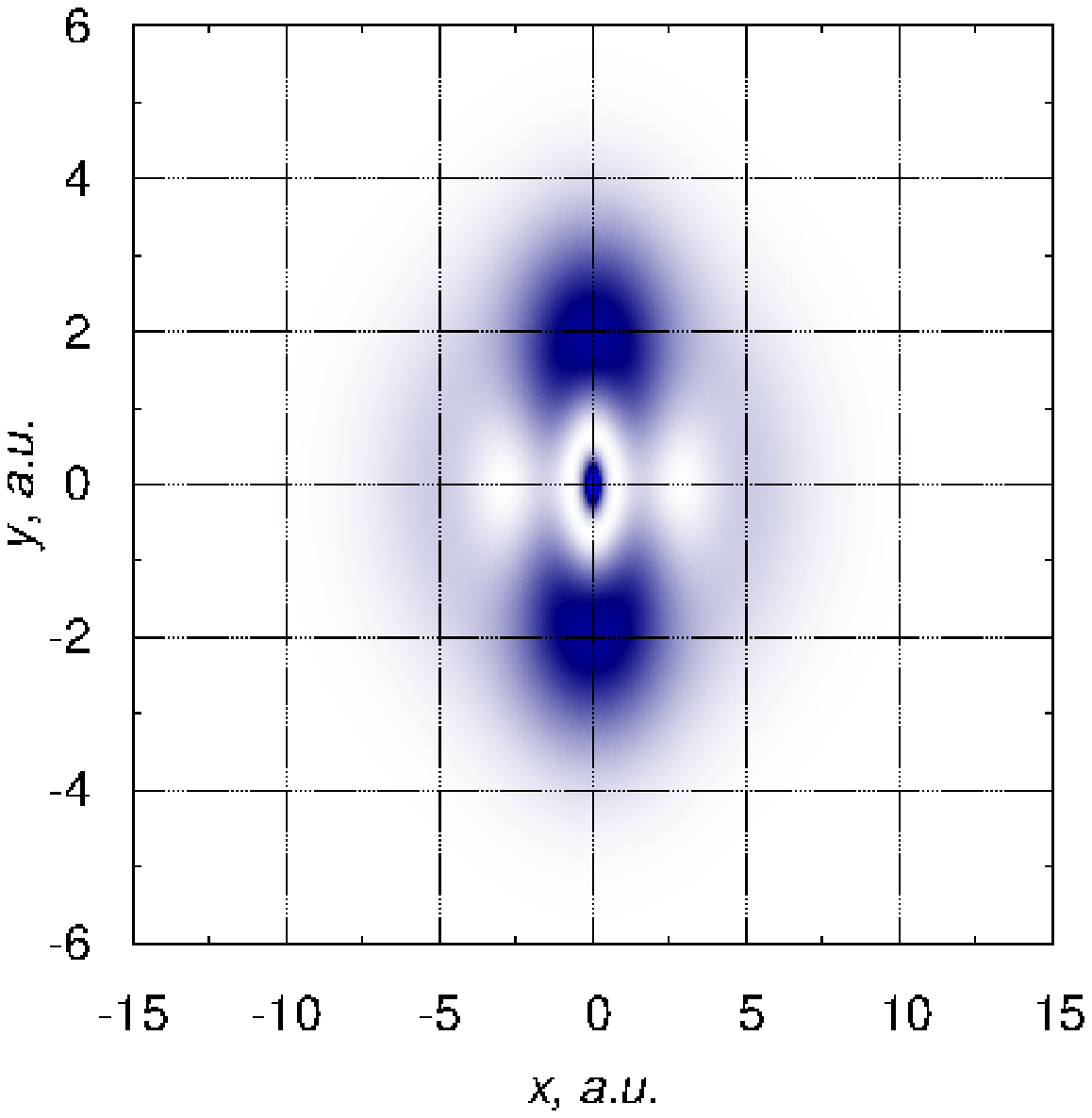}
\put(85,89){\textit{f}}
\end{overpic}
\end{minipage}
\vfill
\hfill
\begin{minipage}[h]{0.3\linewidth}
\begin{overpic}[height=4cm]{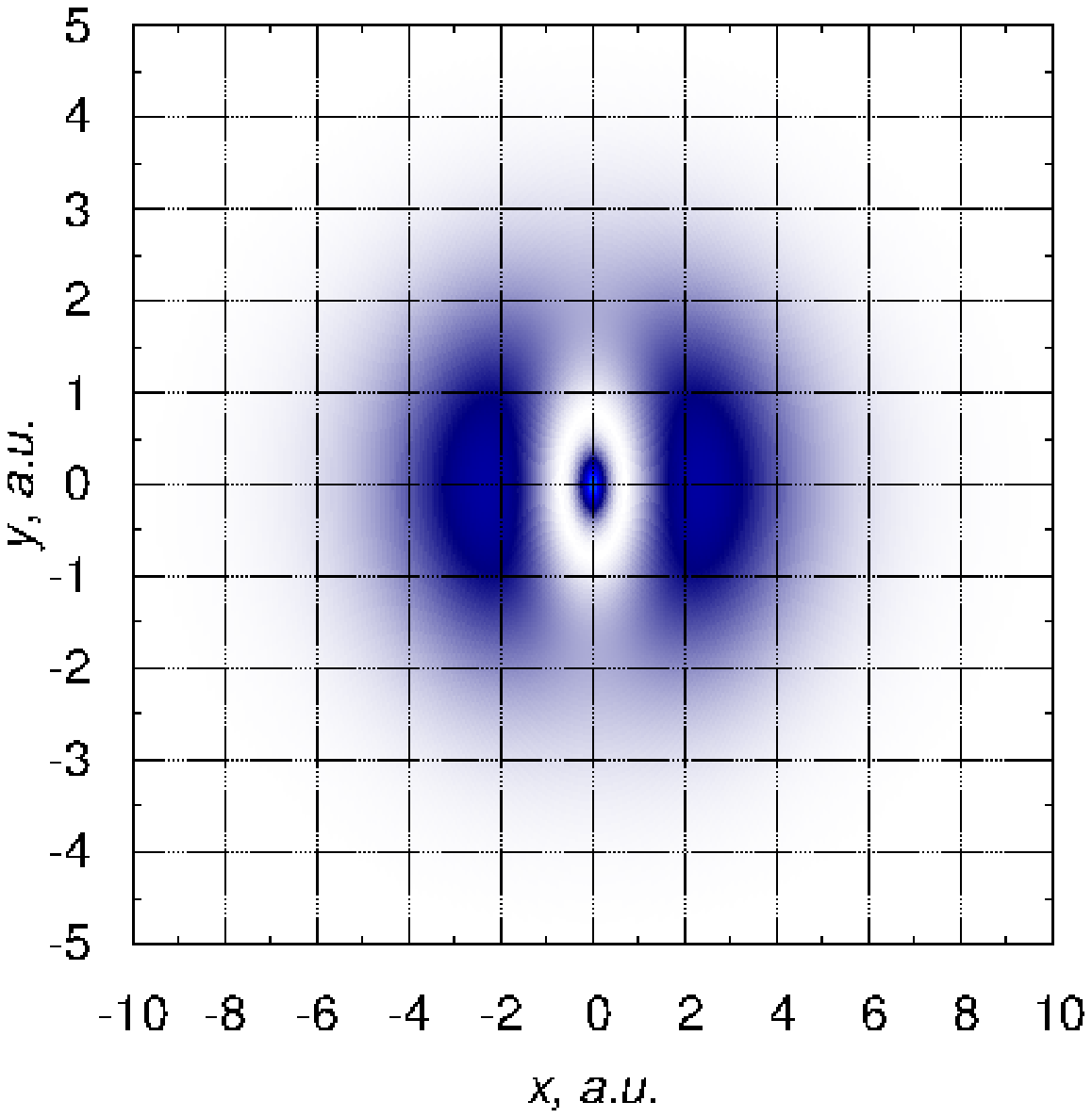}
\put(89,91){\textit{g}}
\end{overpic}
\end{minipage}
\hfill
\begin{minipage}[h]{0.3\linewidth}
\begin{overpic}[height=4cm]{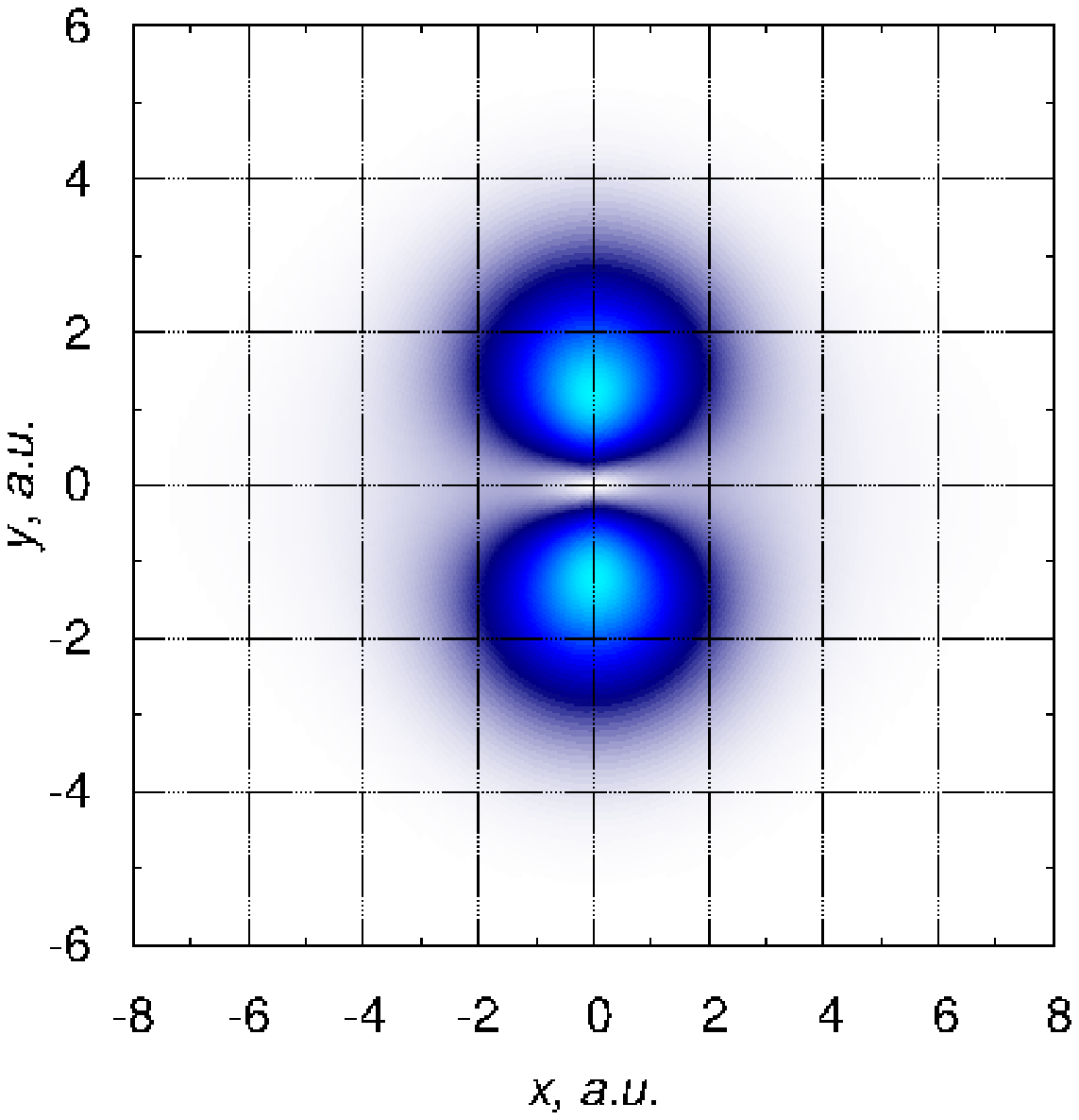}
\put(87,89){\textit{h}}
\end{overpic}
\end{minipage}
\hfill
\begin{minipage}[h]{0.3\linewidth}
\begin{overpic}[height=4cm]{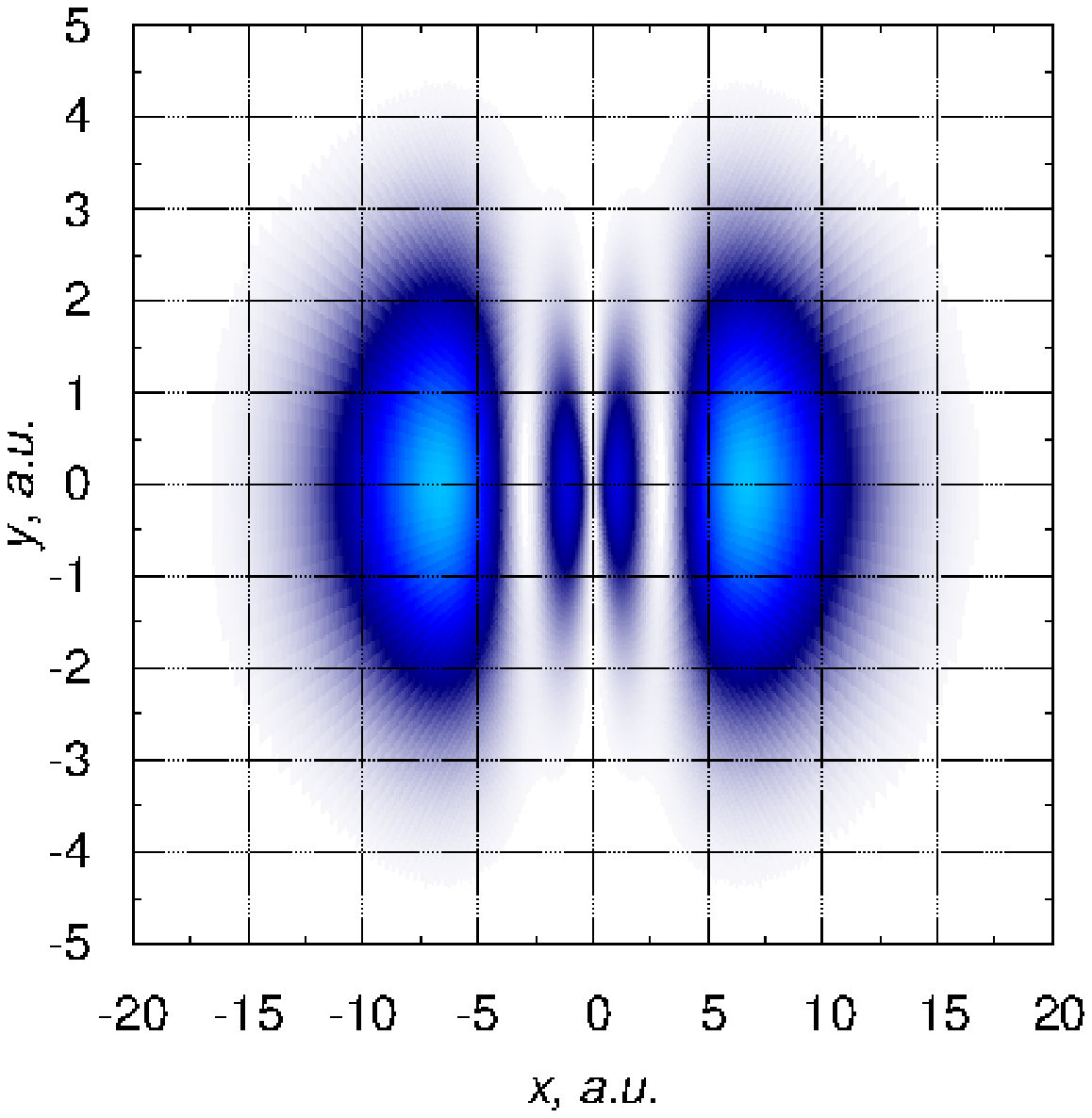}
\put(89,91){\textit{i}}
\end{overpic}
\end{minipage}
\hfill
\caption{
(Color online) Density plot of the square of the absolute value of the wave function $|\Psi(\rho,\phi)|^2$ of the first three ($n=1,2,3$) low-lying 2D hydrogen excited states at $B=0.5$ a.u. at tilt angle
$\alpha=0^{\circ}$ for principal quantum number $n=1$ ($a$),  $n=2$ ($b$),  $n=3$ ($c$); at $\alpha=27^{\circ}$ for $n=1$ ($d$),   $n=2$ ($e$),  $n=3$ ($f$); at $\alpha=54^{\circ}$ for  $n=1$ ($g$),   $n=2$ ($h$),  $n=3$ ($i$); at $\alpha=81^{\circ}$ for  $n=1$ ($j$),  $n=2$ ($k$),  $n=3$ ($l$). Darks blue regions correspond to low and bright blue ones to high densities. Note that $X$ scales are different from $Y$ ones.
}
 \label{figWaveFunctionProfiles}
 \end{figure*}
 
 \newpage
 
\begin{table*}
\begin{ruledtabular}
\begin{tabular}{|c|rrrr|} 
$B$, a.u. & $0^{\circ}$ & $27^{\circ}$ & $54^{\circ}$ & $81^{\circ}$  \\
\hline 
\rule{0mm}{5mm} 
$0.5$ & $-0.27678405$ & $-0.26528554$ & $-0.23126243$ & $-0.17640957$ \\ 
$1$ & $-0.20392330$ & $-0.18813374$  & $-0.14614052$ & $-0.08150442$ \\
$2$ & $0.04130874$ & $0.06072439$ & $0.09548195$ & $0.14038591$ \\
$4$ & $0.67852876$ & $0.69653696$ & $0.67573616$ & $0.62258164$ \\
$10$ & $2.95734799$ & $2.94208198$ & $2.62839018$ & $2.14718594$\\
$100$ & $43.74850351$ & $44.23373675$ & $35.01722779$ & $26.51275203$\\
$1000$ & $480.93444481$ & $456.24798403$ & $371.78493695$ & $277.87663177$\\
$10000$ & $4945.43794585$ & $4650.63170583$ & $3780.99288744$ & $2816.81434104$\\
\end{tabular} 
\end{ruledtabular}
 \caption{
Energy values $E_1$ of the \textbf{first} excited state of 2D hydrogen atom at different magnetic field strengths and tilt angles~$\alpha$.
}
\label{tabE_2}
\end{table*}

\begin{table*}
\label{tabE_3}
\begin{ruledtabular} 
\begin{tabular}{|c|rrrr|} 
$B$, a.u. & $0^{\circ}$ & $27^{\circ}$ & $54^{\circ}$ & $81^{\circ}$  \\ 
\hline 
\rule{0mm}{5mm} 
$0.5$  & $-0.10686554$ & $-0.08765517$ & $-0.07627428$ & $-0.11025339$ \\
$1$    & $ 0.00707007$ & $0.03794284$  & $0.06357988$ & $0.02558182$ \\
$2$    & $ 0.31437038$ & $0.36132446$ & $0.36535958$ & $0.28228436$ \\
$4$    & $ 1.04219925$ & $1.10783242$ & $1.02157341$ & $0.79709702$\\
$10$   & $ 3.50447015$ & $3.57886264$ & $3.12007058$ & $2.39508346$\\
$100$  & $45.41374489$ & $45.60201238$ & $37.19013487$ & $28.16742401$\\
$1000$ & $486.47671280$ & $464.85293413$ & $393.52867231$ & $299.03138826$\\
$10000$ & $4966.59383286$ & $4723.50218673$ & $4022.90347429$ & $3060.77202297$\\
\end{tabular} 
\end{ruledtabular} 
\caption{
Energy values $E_2$ of the \textbf{second} excited state of 2D hydrogen atom at different magnetic field strengths and tilt angles~$\alpha$.
}
\end{table*}

\begin{table*}
\begin{ruledtabular} 
\begin{tabular}{|c|rrrr|} 
$B$, a.u. & $0^{\circ}$ & $27^{\circ}$ & $54^{\circ}$ & $81^{\circ}$  \\ 
\hline 
\rule{0mm}{5mm} 
$0.5$ & $0.04072983$ & $-0.01085554$ &  $0.00438870$ & $-0.05474697$ \\
$1$ & $0.49549097$ & $0.1425023882$  & $0.17184677$ & $0.15375924$ \\
$2$ & $1.57853526$ & $0.50673360$ & $0.50625030$ & $0.42459900$ \\
$4$ & $4.00329697$ & $1.31189496$ & $1.22852674$ & $1.00456007$\\
$10$ & $11.89181281$ & $3.90040567$ & $3.54926952$ & $2.86808926$\\
$100$ & $140.5148315$ & $46.95758876$ & $41.68141369$ & $33.12068670$\\
$1000$ & $1470.80545235$ & $477.30100986$ & $438.29023420$ & $345.22282692$\\
$10000$ & $14854.71822028$ & $4847.91395473$ & $4451.92041328$ & $3496.02256291$\\
\end{tabular} 
\end{ruledtabular} 
\caption{
Energy values $E_3$ of the \textbf{third} excited state of 2D hydrogen atom in at different magnetic field strengths and tilt angles~$\alpha$.}
\label{tabE_4}
\end{table*}

\twocolumngrid


\begin{thebibliography}{33} 

\bibitem{Soylu_2006} A.~Soylu, O.~Bayrak and I.~Boztosun, Int. J. Mod. Phys. E \textbf{15}, 1263 (2006).

\bibitem{Turbiner_2014} M.~A.~Escobar-Ruiz and A.~V.~Turbiner, Ann. Phys. (N. Y.) \textbf{340}, 37 (2014).

\bibitem{Turbiner_2015} M.~A.~Escobar-Ruiz and A.~V.~Turbiner, Ann. Phys. (N. Y.) \textbf{359}, 405 (2015). 

\bibitem{Chen_1991} R.~Chen, J.~P.~Cheng, D.~L.~Lin et al., Phys. Rev. B \textbf{44}, 8315 (1991).

\bibitem{Villalba_1996} V.~M.~Villalba and R.~Pino, J. Phys.: Condens. Matter \textbf{8}, 8067 (1996).

\bibitem{Soylu_2007} A.~Soylu and I.~Boztosun, Physica B: Condens. Matter \textbf{396}, 150 (2007).

\bibitem{Kallin_1984} C.~Kallin, and B.~I.~Halperin, Phys. Rev. B, {\bf 30}, 5655 (1984).

\bibitem{Portnoi_2002} D.~G.~W.~Parfitt and M.~E.~Portnoi, J. Math. Phys. \textbf{43}, 4681 (2002).

\bibitem{Gutzwiller_1971} M.\,C.~Gutzwiller J. Math. Phys. {\bf 12}, 343 (1971)

\bibitem{Friedrich_1989} H.~Friedrich and H.~Wintgen, Phys. Rep., {\bf 183}(2), 37-79 (1989).

\bibitem{Harada_1983} A.~Harada and H.~Hasegawa, J. Phys. A: Math. Gen. {\bf 16} 259 (1983).


\bibitem{Gutzwiller_2013} M.\,C.~Gutzwiller, \textit{Chaos in classical and quantum mechanics}, Vol. 1 (Springer Science \& Business Media, 2013).

\bibitem{Bohigas_1984}  O. Bohigas and M. J. Giannoni, in \textit{Mathematical and
Computational Methods in Nuclear Physics}, Lecture Notes in Physics, Vol. 209, edited by J. S. Dehesa et al. (Springer, Berlin, 1984).


\bibitem{Delande_1986} D.~Delande and J.\,C.~Gay Phys. Rev. Lett., {\bf 57}, 2006 (1986).

\bibitem{Monteiro_1990} T.\,S.~Monteiro and G.~Wunner, Phys. Rev. Lett. {\bf 65}, 1100 (1990).

\bibitem{Grabowski_1994} P.~Hawrylak, and M.~Grabowski, Phys. Rev. B {\bf 49}, 8174 (1994).

\bibitem{Melezhik_1991} V.~S.~Melezhik, J. Comput. Phys. \textbf{92}, 67 (1991).

\bibitem{Melezhik_2003} P.~Capel, V.~S.~Melezhik and D.~Baye, Phys. Rev. C \textbf{68}, 014612 (2003).

\bibitem{Koval_2016} E.\,A.~Koval and O.\,A.~Koval, Zh. Eksperim. i. Teor. Fiz. \textbf{152} (2017) [In print, in Russian]; ArXiv quant-ph/1701.06235.





\bibitem{Haake_2013} F.~Haake, \textit{Quantum signatures of chaos}, Vol. 54 (Springer Science \& Business Media, 2013).

\bibitem{Gelfand_2000} I.~M.~Gelfand and S.~V.~Fomin, \textit{Calculus of Variations}, Dover Publications, New York, (2000).

\bibitem{Robnik_2003} M.~Robnik and V.\,G.~Romanovski, J. Phys. A: Math. Gen. \textbf{36}, 7923 (2003).

\bibitem{Butov_2001} L.\,V. Butov, C.\,W. Lai, D.\,S. Chemla, Y.\,E. Lozovik, K.\,L. Campman and A.\,C. Gossard, Phys. Rev. lett. \textbf{87}, 216804 (2001).

\bibitem{Lozovik_2002} Yu.\,E. Lozovik, I.\,V. Ovchinnikov, S.\,Yu. Volkov, L.\,V. Butov and D.\,S. Chemla, Phys. Rev. B \textbf{65}, 235304 (2002).


\end{thebibliography}
\end{document}